\documentclass[12pt,preprint]{aastex}
\usepackage{graphicx}

\def\msun{\ifmmode \rm{M}_\odot \else M$_\odot$\fi}     
\def\mdot{\ifmmode \dot M \else $\dot M$\fi}   	        
\def\about{\ifmmode \sim \else $\sim$ \fi}		
\def\c2{\ifmmode \chi^2/\nu \else $\chi^2/\nu$ \fi}	

\begin{document}
\title{X-RAY OBSERVATIONS OF V4641 SGR (= SAX J1819.3-2525) DURING THE BRIEF AND VIOLENT OUTBURST OF 2003}

\author{Dipankar Maitra and Charles D. Bailyn}
\affil{Yale University, Department of Astronomy, P.O.Box 208101, New Haven, CT, 06520-8101}
\email{maitra,bailyn@astro.yale.edu}

\begin{abstract}
We present the results of detailed analysis of pointed X-ray observations by 
RXTE PCA/HEXTE 
of the black hole X-ray binary (BHXRB) system V4641 Sgr (= SAX J1819.3-2525) 
during its outburst of August 2003. Soft X-ray (3-20 keV) flux variations by 
factors of 10 or more on timescales of minutes or shorter were seen. The 
rapid and strong variability of this source sets it apart from typical XRBs. 
In spite of large luminosity fluctuations, the spectral state of the source 
did not 
change significantly during the dwells which suggests that the physical 
emission processes did not change much during the observations. The energy 
spectra during the dwells were dominated by a hard Comptonized powerlaw 
component, indicative of the canonical low/hard state observed in other BHXRBs.
No soft thermal component was found in three out of the four RXTE pointings. 
However spectral deconvolution of the observation with largest average 
luminosity suggests an obscured, hot accretion disk. During one of the 
observations we detected a short term (\about 100s) soft X-ray dropout which is
apparently due to variability in the observed column density. Strong Fe 
K$\alpha$ fluorescent emisssion line near 6.5 keV was detected with large 
equivalent widths in the range of 700 - 1000eV. In the temporal domain, the 
Fourier power spectra were dominated by red noise below a few Hz. Poisson noise
dominated at higher frequencies and no high frequency features were detected. 
The strong Comptonized spectra, broad iron emission line, absence of disk 
component in the spectra, absence of any timing variability above few Hz and  
occasional large changes in the column density along the line-of-sight, all 
support an enshrouded black hole with occasional outflow and a very dynamic 
environment.
\end{abstract}


\section{Introduction\label{sec:Intro}}

Outbursts in transient X-ray binary systems typically span orders of magnitude
in luminosity and mass accretion rate. This allows one to study accretion 
physics over a wide range as well as general relativity in the strong field 
limit. 
During a complete outburst cycle, the sources exhibit a multitude of states 
that have very different spectral and temporal characteristics from each other.
Two canonical states that most transients seem to go through are (i) a hard 
powerlaw dominated state with high rms variability in the lightcurve, termed 
the {\em low/hard} state and (ii) the {\em high/soft} or {\em thermal dominant}
state with soft photon dominated spectra from the accretion disk and little 
rms variability. Sometime a very high state with steep powerlaw is also seen
(see \citet{tl1995} and \citet{mr2005} for a recent review).

V4641 Sagitarii (SAX J1819.3-2525) was discovered by \citet{g1978} when it went
through an optical flare and hence it was categorized as an irregular optical 
variable. Initially this system was confused with another neighboring 
long-period variable called GM Sgr, therefore many old literature erroneously 
refer to GM Sgr as the optical counterpart of the strong X-ray source. The 
system was first detected in X-ray wavelengths in February 1999, independently 
by {\em BeppoSAX} \citep{i1999} and RXTE \citep{m1999} and designated as 
SAX J1819.3-2525. V4641 Sgr is presently known to be a compact binary system 
harboring a $9.61^{+2.08}_{-0.88} \msun$ black hole candidate and a 
$6.53^{+1.6}_{-1.03} \msun$ B9 secondary star, with an orbital period of 
$2.81730\pm0.00001$ days \citep{o2001}. \citet{o2001} estimated the distance to
the system between 7.4 to 12.3 kpc and an orbital inclination between 
$60\degr-70\degr$. We have used an inclination of $65\degr$ and a distance of 
$10$ kpc for all the calculations in this paper.

During the outburst of V4641 Sgr in 1999, RXTE observations showed flaring 
X-ray activity with Super-Eddington luminosity \citep{w2000,r2002} on Sep 15. 
Rapid variability in optical brightness were also reported \citep{s1999,u2002} 
during this outburst. Radio observations during the outburst revealed a 
marginally resolved structure \citep{h2000a}. The inferred superluminal proper 
motions of the radio structure ($\gtrsim0.22$ arcseconds per day) were 
attributed to relativistic motion of a radio jet. This led to its 
classification as a possible galactic microquasar like GRS 1915+105 
\citep{m1994} and GRO 1655-40 \citep{t1995,h1995}. Based on the X-ray 
observations, \citet{r2002} suggested the formation of an extended 
envelope/outflow around the source during the outburst. Optical spectroscopy 
by \citet{c1999} also suggest the presence of strong wind. 

Since the 1999 outburst, activity from the source has been reported in 2000
\citep{h2000b}, 2002 \citep{u2004}, 2003 \citep{atel170} and 2004 \citep{s2004}
In all cases, the entire span of the outburst cycles for this source is much 
shorter than that of typical compact transient systems. Also, V4641 Sgr does 
not exhibit a typical {\em Fast Rise and Exponential Decay} (FRED) 
\citep{csl1997} lightcurve profile during the outburst. 

In 2003, signs of activity was first noted by 
VSNET\footnote{http://ooruri.kusastro.kyoto-u.ac.jp/mailman/listinfo/vsnet-alert} group on Aug 01, 2003 
and shortly thereafter by the {\em Small and Moderate Aperture Research 
Telescope System} (SMARTS)\footnote{http://www.astro.yale.edu/smarts/} 
consortium telescopes at Cerro Tololo Inter-American Observatory (CTIO) in 
Chile \citep{atel170}. Subsequent multiwavelength observations showed that the 
source was active in X-rays \citep{atel171} and radio \citep{atel172} as well.
As is characteristic for this enigmatic source, rapid time variability was 
observed in data obtained from radio, optical as well as X-rays. The observed 
X-ray spectrum was very hard in nature, i.e. a strong contribution of hard 
X-rays were seen compared to the soft X-ray flux. The source of the hard X-ray 
radiation is believed to be either inverse Compton scattering of soft thermal 
photons from the accretion disk \citep{st1980} and/or synchrotron emission from
a jet \citep{mff2001}. Broad and strong iron fluorescent emission lines near 
6.5 keV were observed. These lines are thought to be produced by reprocessing 
of hard X-rays impinging on cold matter, usually the accretion disk 
\citep{gf1991,rn2003}, where principal factors that causes the line to broaden 
are strong gravitational field near the compact source and relativistic motion 
of the radiation emitting particles (relaticistic Doppler broadening). The line
may also be created in a plasma (or corona) close to the central black hole. 
The line shape in this case is intrinsically Gaussian, the line energy depends 
on the dominant ionization states and the line width depends on rotation and 
Compton scattering in the corona \citep{kw1989}.

In this paper we report the detailed spectral and temporal analyses of pointed 
RXTE observations of the 2003 August outburst of V4641 Sgr. In 
\S\ref{sec:Observations} we describe the general data reduction procedures 
adopted in this paper. In \S\ref{sec:CC}, \S\ref{sec:Spectroscopy} and 
\S\ref{sec:Timing} the results of the color evolution, spectral and timing 
analyses of the data are presented respectively. The conclusions are 
summarised in \S\ref{sec:Summary}.

\section{OBSERVATIONS AND RESULTS\label{sec:Observations}}

\begin{figure}
\begin{center}
\includegraphics[height=3.0in]{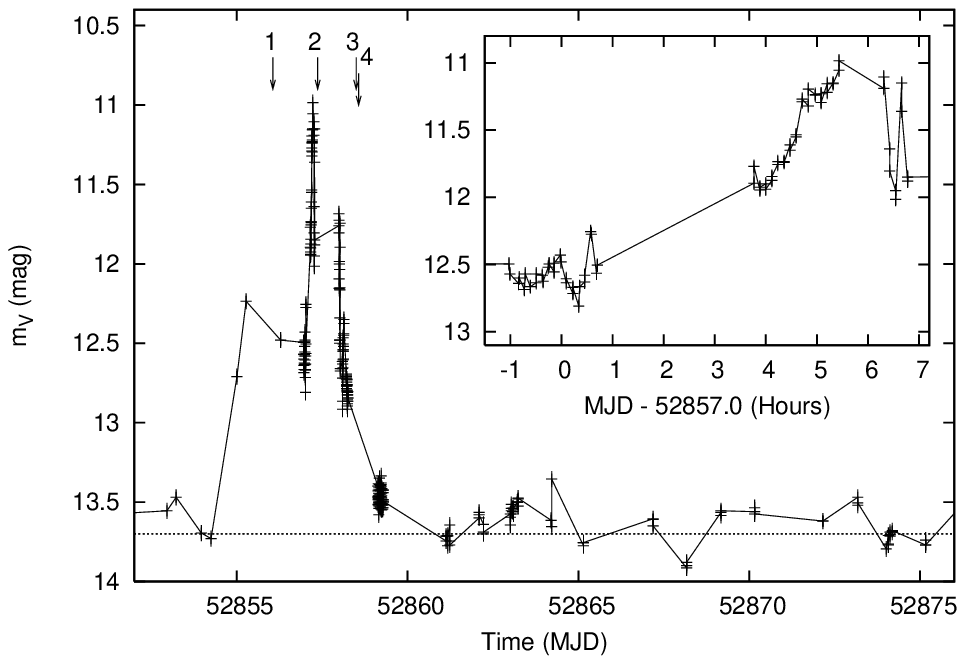}
\includegraphics[height=3.0in]{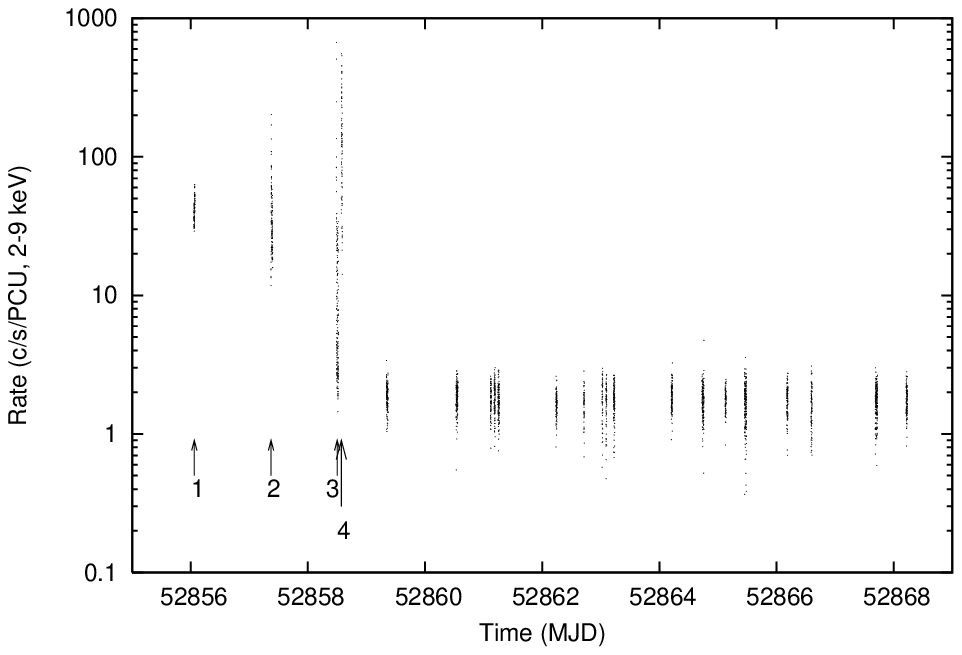}
\caption{Top: V band lightcurve of V4641 Sgr during the Aug 2003 outburst. 
Four vertical lines on the top labelled 1-4 are the times when significant 
X-ray activity was observed by RXTE (see Table~\ref{tab:obslog}). The 
horizontal dotted line represents the mean quiescent brightness of 13.7 
magnitude. 
The inset is a zoom near MJD 52857.0 to show intra-night optical variability.
Bottom: X-ray activity of V4641 Sgr during the Aug 2003 outburst as seen by
RXTE. The count-rates presented are 2-9 keV background subtracted rates. The 
four vertical lines labelled 1-4 near the bottom are the observation dwells
that have been studied in this paper (see Table~\ref{tab:obslog}). The PCA 
background level is about 2 millicrab or 5.57 counts/PCU. The optical and X-ray 
observations shown here are not coincident. Simultaneous optical observations 
were obtained during dwell (4) using the NCU Lu-Lin Observatory, Taiwan and 
reported separetely \citep{m2004,b2005}.\label{outburst}}
\end{center}
\end{figure}

We used optical observations of V4641 Sgr from the SMARTS consortium 1.3m 
telescope to trigger RXTE target of opportunity observations. 
The optical data were obtained using the ANDICAM (A Novel Dual Imaging 
CAMera, see \citet{d2003} for details)
instrument mounted on the SMARTS 1.3m telescope at Cerro Tololo 
Inter-American Observatory. The ANDICAM detector consists of a dual-channel 
camera that allows for simultaneous optical and IR imaging.
The V band lightcurve from MJD 52840 to MJD 52870 (July 20, 2003 - Aug 19, 
2003) in Fig.~\ref{outburst} shows rapid variability and short duration of 
the entire outburst, both of which are characteristic of previous outbursts 
of this source\citep{u2004,w2000}.

RXTE \citep{brs1993} pointed observations of V4641 started on MJD 52856 and 
continued till MJD 52868 (2003 Aug 5 - 2003 Aug 17). The observable X-ray 
activity ceased after MJD 52858 (Aug 7, see Fig.\ref{outburst}). 
Between Aug 5 and Aug 7, four RXTE pointed observations were done. A list of 
observation start times and durations is given in Table~\ref{tab:obslog}.

{\em HEASOFT FTOOLS} (v5.3) software was used to perform the X-ray data 
reduction. PCA {\em Science Array} data-mode with 16s time bins were used for
color analysis and  
spectral data extraction. We extracted the lightcurves and spectral information
from the top layer of the second proportional counter unit (PCU2) using the 
{\em Standard 2} files. 
Standard screening criteria as described in the RXTE {\em Cookbook\footnote
{http://rxte.gsfc.nasa.gov/docs/xte/recipes/cook\_book.html}} were applied to 
select segments of good data when (1) the source was at least 10 degrees away 
from 
earth's limb and (2) the satellite was not passing through the region of 
South Atlantic Anomaly and (3) the pointing was stable with pointing offsets 
less than 0.02 degrees. Since the source was not very bright, we also excluded
regions with significant electron contamination (ELECTRON2 $<$ 0.1). The errors 
in count-rates and hardnesses are $1\sigma$ errors assuming a Poissonian 
distribution for the recorded photon count-rates. 
Background spectra were estimated from the latest models 
\footnote{http://heasarc.gsfc.nasa.gov/docs/xte/pca\_news.html\#quick\_table}
 using {\em pcabackest} (v3.0) tool. {\em Pcarsp} (v8.0) was used to generate 
the redistribution matrix files (rmf) and ancilliary response files (arf) and 
then combined to form a single response file (rsp). We used the HEXTE 
{\em Archive} 
mode data from cluster A with 64 energy channels for total energy band 15-250 
keV with a timing resolution of 16s. While fitting simultaneous PCA+HEXTE 
spectra, the normalization of the HEXTE data were left as a free parameter to 
match the PCA spectra. 

Temporal information was extracted either from the {\em Science Event} files 
with $2^{-13}\mu$s ($\about 122\mu$s) time resolution ({\em DATAMODE = 
E\_125us\_64M\_0\_1s}) or from {\em GoodXenon} modes of time resolution 
$2^{-20}\mu$s ($\about 0.95\mu$s), whichever available. However for the third 
pointing, we used the 
{\em E\_16us\_16B\_36\_1s} datamode which covers almost the entire PCA energy 
spectrum from 14.9 keV and above with a timing resolution of $2^{-16}\mu$s 
($\about 15.2\mu$s) and 
readout time of 1s, since no event data with full PCA energy range were taken 
for this dwell. Data from all the detectors and their layers were combined. 
The event data for all the dwells were binned to $2^{-13}\mu$s
time bins (corresponding to a Nyquist frequency of 4096 Hz) to ensure uniform 
analysis. Temporal studies were done in 
the Fourier domain using the {\em powspec} (v1.0) tool to carry out the 
Fast-Fourier Transforms and create the power density spectra (PDS) following 
the prescription of \citet{vdk1989,vdk1995}. Since the count-rates were
not high, the dead-time effect is negligible and we did not explicitly correct
the light curves for dead time. Instead we modelled the dead-time modified 
Poisson noise in the PDS with a constant and subtracted this constant to 
calculate the Fourier power. The PDS were normalized w.r.t count-rate so that 
the resulting power spectra are in the units of (rms/mean)$^2$/Hz. The PDS were
then logarithmically rebinned in frequency space to reduce noise at high 
frequencies. Unless otherwise stated, 
the error bars in the average power spectra were calculated by evaluating the 
standard deviation of the average power for each frequency. 
However the error bars for the individual 64s power spectra during dwell (2) 
were calculated by propagating through theoretical error bars obtained 
from the relevant chi-square distribution. The errors in 
spectral and temporal model fits are 90\% confidence regions for a single 
parameter ($\Delta\chi^2=2.706$).

\subsection{Lightcurve and Colors\label{sec:CC}}
\begin{figure}
\begin{center}
\includegraphics[height=3.2in, angle=-90]{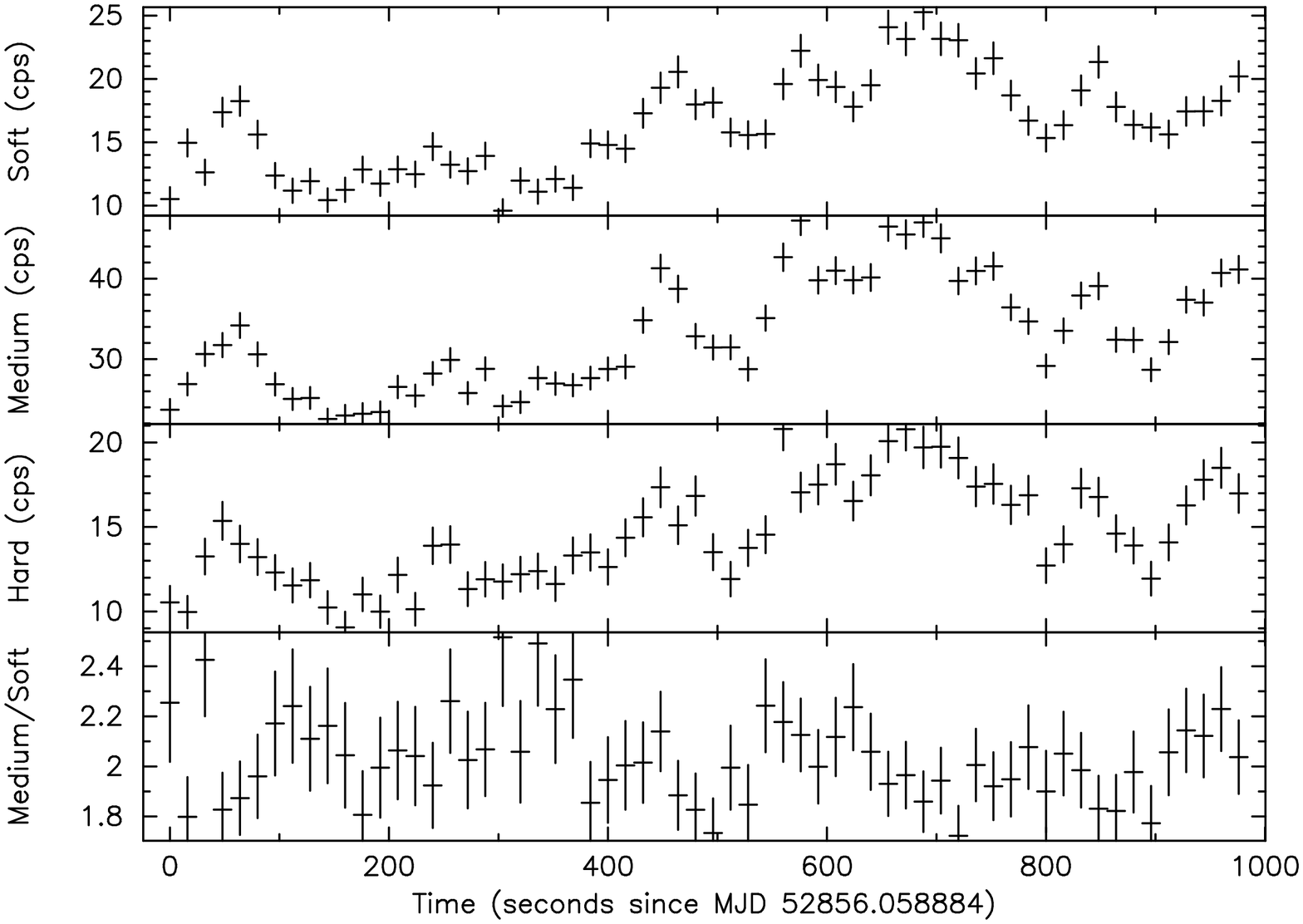}
\includegraphics[height=3.2in, angle=-90]{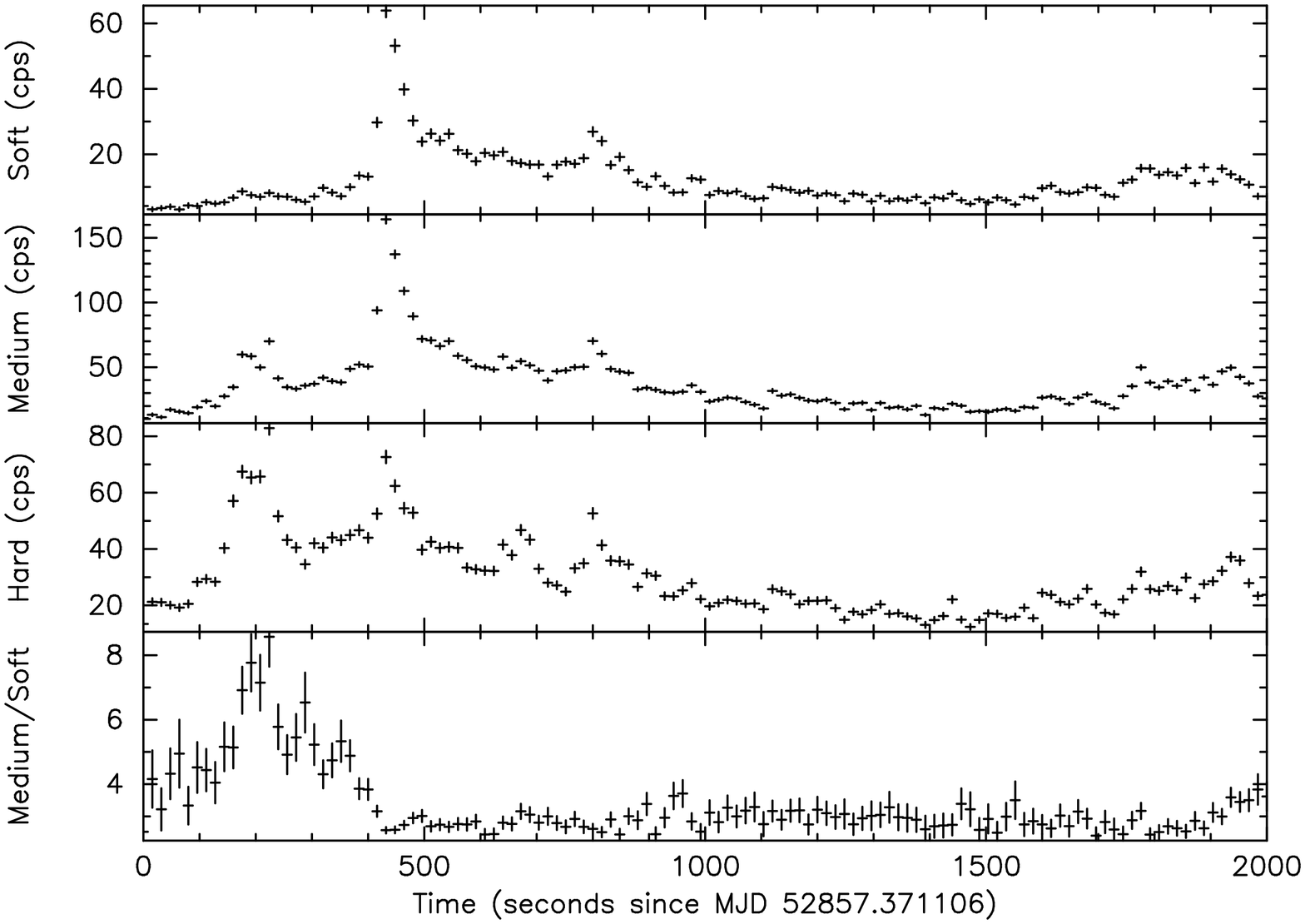}
\includegraphics[height=3.2in, angle=-90]{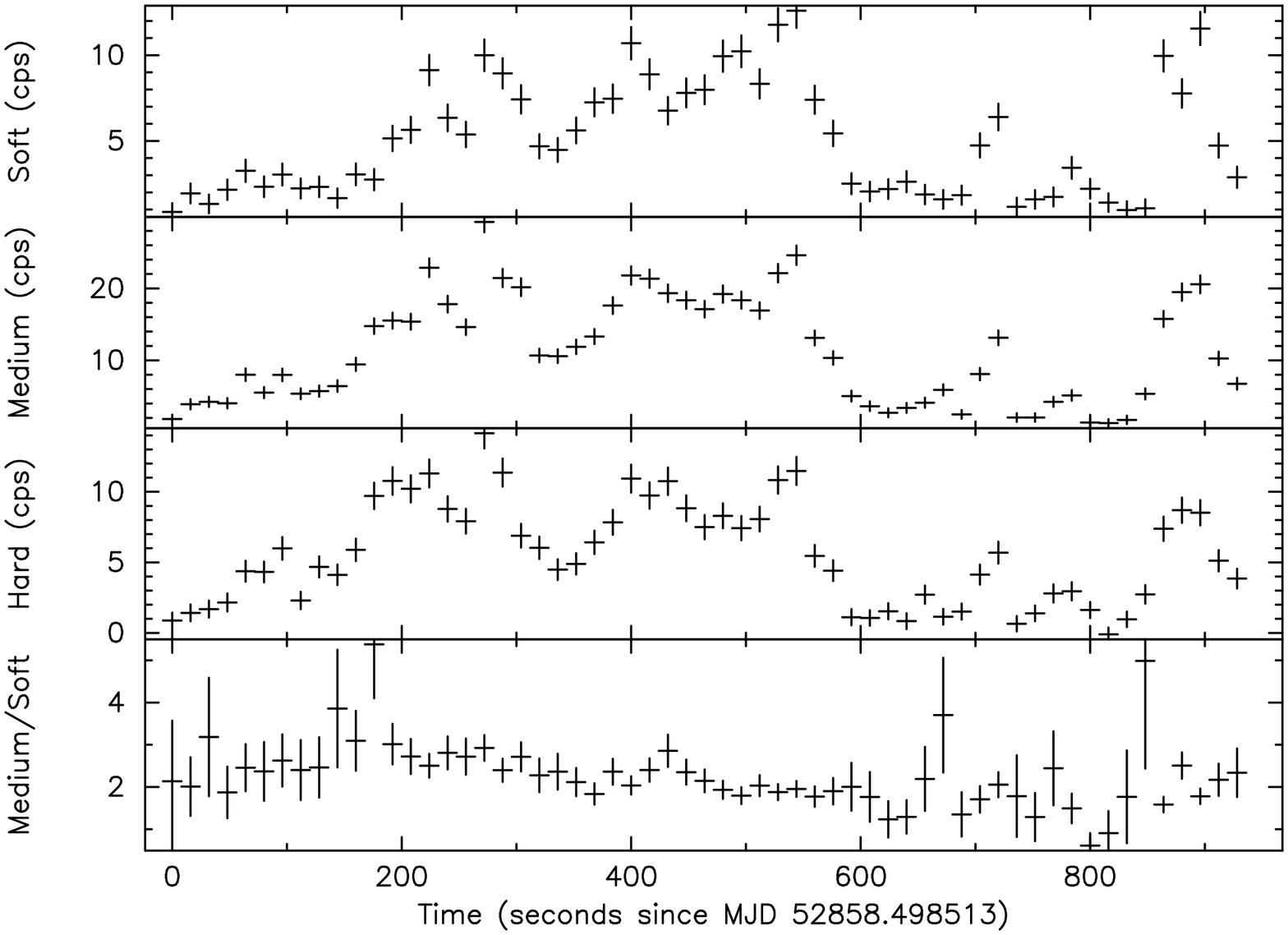}
\includegraphics[height=3.2in, angle=-90]{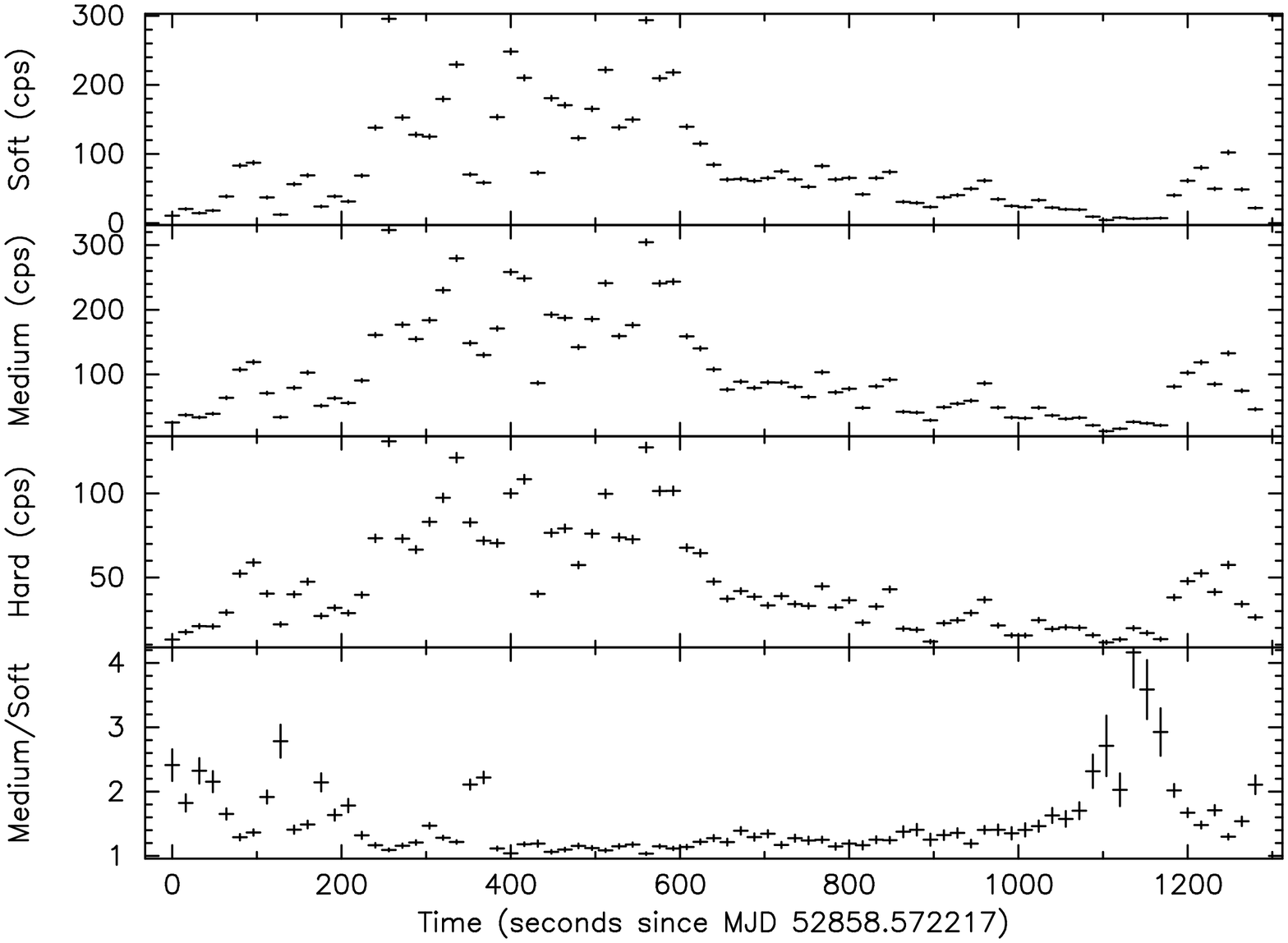}
\caption{X-ray flux and hardness ratio variations with time for the observed 
RXTE pointings. Dwell (1) is shown on top left, dwell (2) on top right, dwell
(3) on bottom left and dwell (4) on bottom right. Each dwell is subdivided in 
four subpanels where the top subpanels show 2.0-5.3 keV PCU2 count-rates (Soft),
the second subpanels show 5.3-10.3 keV PCU2 count-rates (Medium), third shows
the 10.3-20.4 keV PCU2 count-rates (Hard) and the fourth subpanels show the 
Medium/Soft hardness ratio.
\label{lc_hr}}
\end{center}
\end{figure}

The PCA lightcurves and (5.3-10.3 keV)/(2.0-5.3 keV) color for the observed 
dwells are shown in Fig.~\ref{lc_hr}. Here we see the highly variable nature
of this source. Intensity variations by an order of magnitude in 
timescales as short as minutes are seen. In dwell (3) the observed count-rates
were the lowest of the four dwells presented in this paper. In contrast, the 
immediately succeeding pointing, viz. the fourth dwell, shows the highest 
count-rates. While the starting time of the fourth dwell is only 1.5 
hours after the ending time of the third dwell, the average count-rate of the 
source increased by a factor of 8.4 compared to dwell (3) and the peak 
count-rate is 14.4 times larger than that of dwell (3). During the fourth 
dwell, simultaneous optical observations of the source were done from Lu-Lin 
Observatory in Taiwan which showed strong X-ray/optical correlation and delayed
arrival of the optical light \citep{b2005,m2004}. Also, of all the RXTE 
pointings during this outburst, this is the last dwell where we observed any 
activity from the source (see Fig~\ref{outburst}).

\begin{figure}
\begin{center}
\includegraphics[height=4.0in, angle=0]{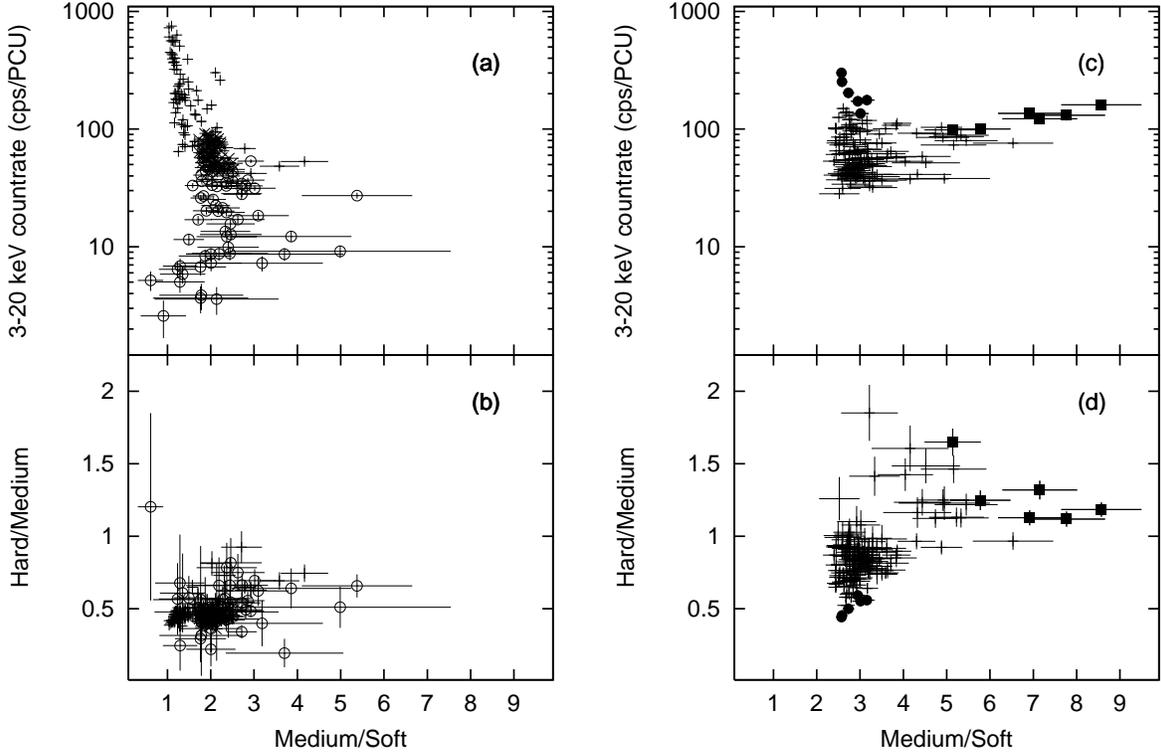}
\caption{Panel (a): Medium/soft band hardness ratio is plotted against the 
3-20keV background subtracted count-rates for dwells (1),(3) and (4). 
The corresponding color-color plot is shown in panel (b).
In both 
plots, data from dwell (1) are shown by crosses ($\times$), data from dwell (3)
are shown by open circles ($\circ$) and the data from dwell (4) are shown by 
plusses (+). In panels (c) and (d) we show the same plots as (a) and (b) 
respectively, but just for dwell (2) where we observed some violent hard and 
soft flaring activities. The data taken during the hard flare are shown by 
filled squares ($\blacksquare$), 
those during the soft flare are shown by filled circles ($\bullet$) and the 
data taken during the rest of the dwell are shown by plusses (+).\label{hd-col}}
\end{center}
\end{figure}

Contrary to what is generally observed in XRBs, the count-rates 
in the medium energy band (5.3-10.3 keV) are greater than that of 
the soft energy band (2.0-5.3 keV). This is due to the extremely strong and 
wide Iron emission line flux near 6.5 keV. As shown in Fig.~\ref{hd-col}(left 
panel), the 
hardness ratios (colors) do not evolve significantly during dwells (1), (3) and
(4), suggesting that the overall spectral state of the source did not change 
during these dwells. There is a slight softening at highest luminosities during
the fourth dwell which is due to the appearance of an otherwise obscured soft 
disc photons and is discussed in detail in \S\ref{sec:Spectroscopy}. 
As seen in the lightcurves, the second dwell is highly dynamic in character. 
The hardness-time and
color-color plots show that the spectral characteristics of the two flares are 
intrinsically very different. While the first flare near 200s is very hard in 
nature, the second flare near 450s is predominantly soft. The two flares occupy 
different regions of the color-color space, whereas the non-flaring state is 
distinct from either of the flaring states. There seems to be no 
major spectral state changes after the second flare. There is some 
overlap between the soft, hard as well as non-flaring state in color-luminosity
space. The non-flaring state in this dwell is spectrally somewhat harder than 
that of the other dwells.
 
\subsection{Spectroscopy\label{sec:Spectroscopy}}

\begin{figure}
\begin{center}
\includegraphics[height=6.0in, angle=-90]{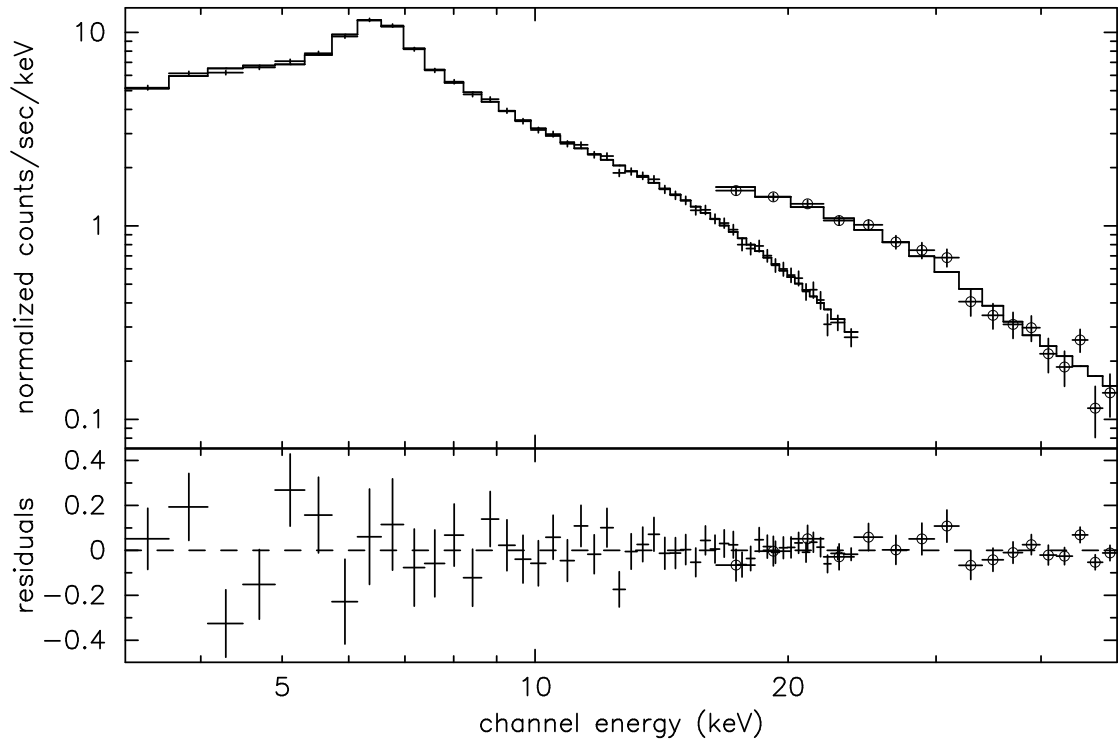}
\caption{Observed count rate PCA (+) and HEXTE ($\oplus$) spectrum and 
Comptonized spectrum model (histogram) for the average emission during dwell
(1) is shown on the top panel and the residuals in the bottom panel. Note the 
strong Fe line feature between 6 and 7 keV.
\label{d1-espectrum}}
\end{center}
\end{figure}

\begin{figure}
\begin{center}
\includegraphics[height=6.0in, angle=0]{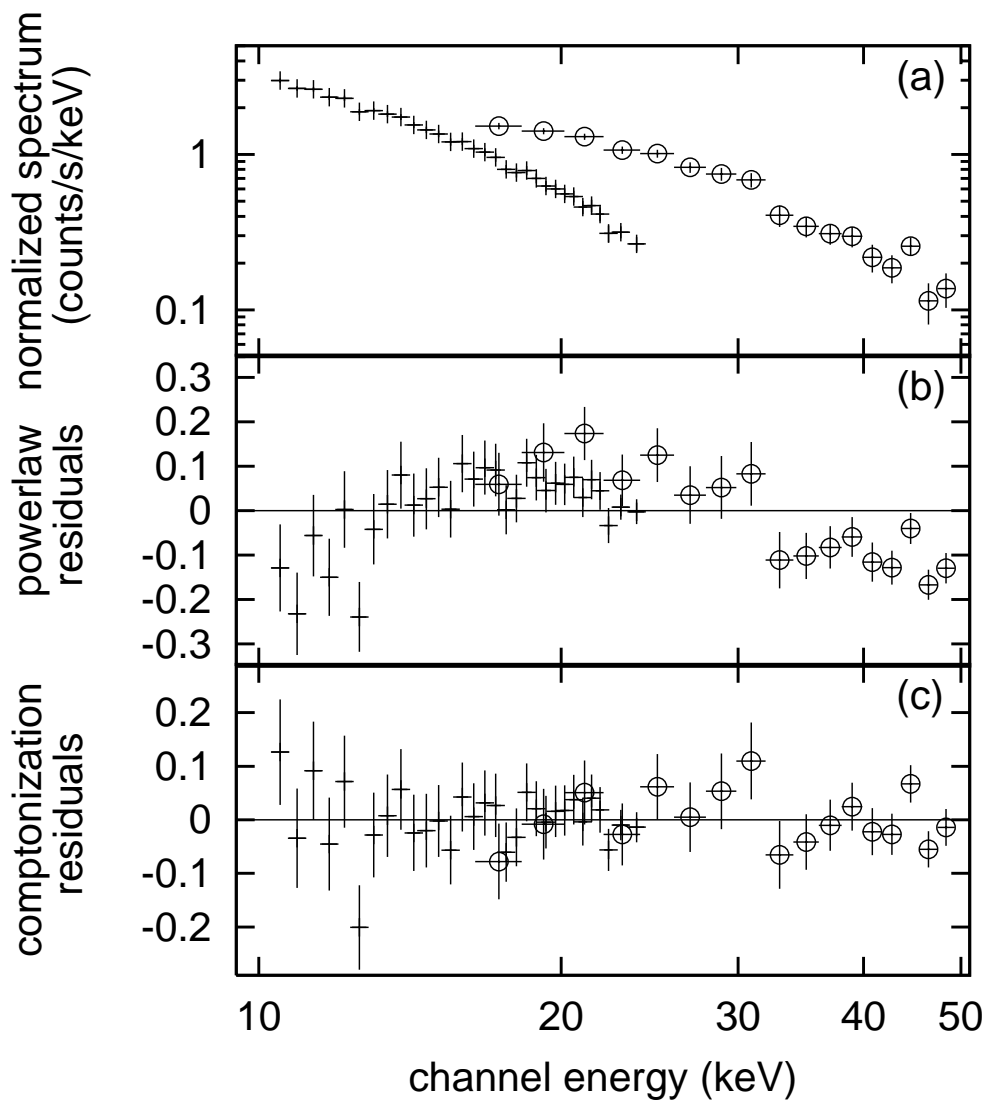}
\caption{
Panel (a): Time averaged counts spectrum during dwell (1).The PCA data are 
marked by $+$ and HEXTE data by $\oplus$ symbols.
Panel (b): Residuals to a powerlaw fit to the energy spectrum in panel (a). 
Systematic deviations from a powerlaw is evident in this panel.
Panel (c): Residuals to a Comptonized powerlaw fit to the same data. There 
is no systematic deviation, although the sharp change just above 30 
keV seen in the HEXTE spectrum is still unaccounted for.
\label{d1-high-espectrum}}
\end{center}
\end{figure}

Since the spectral state of the source did not change appreciably during 
dwells (1), (3) and (4), we analyzed time averaged spectral and temporal 
properties of each of these dwells. The most notable feature in the combined 
spectrum is the strong, broad emission line complex around 6.5 keV which is 
easily seen in the counts spectrum in Fig.~\ref{d1-espectrum}. Since such a 
strong line makes 
determination of continuum parameters difficult, we initially modelled the 
spectral region harder than 10 keV, with the assumption that this region is 
free from strong emission/absorption features or edges. For most BHXRB systems, 
these energy ranges are dominated by powerlaw photons. However, in this case 
we found that a simple powerlaw model fails to describe the 
spectrum (e.g. $\chi^2/\nu = 140/45$). An appreciable 
amount of curvature was 
observed in the residuals, shown in Fig.~\ref{d1-high-espectrum}(a), which is 
usually taken to be a reflection effect. We therefore used
the {\em pexrav} model by \citet{mz1995} which calculates an exponentially cut 
off power law spectrum reflected from neutral material. This model provides a 
much better fit to the data with $\chi^2/\nu = 35/44$ as shown in 
Fig.~\ref{d1-high-espectrum}(b). However, the scaling factor for reflection is 
not well constrained by the fit and was therefore fixed to 1.0 (corresponding 
to an isotropic source above the disk). The overall PCA and HEXTE 3-50 keV 
spectrum 
was modelled using a warm photoelectric absorber \citep{mm1983} of hydrogen 
column density fixed to $2.3\times10^{21}$ atoms/cm$^2$ \citep{dl1990} and the 
Comptonized powerlaw. The abundances were taken from the table by 
\citet{ag1989}.
Modelling the Fe line complex for dwell (1)
required two Gaussians, most likely corresponding to different ionization 
states of iron. The moderate energy resolution of the 
spectrometer makes it difficult to determine the exact ionization states of 
these lines.
We tried the relativistic line model by \citet{l1991} to fit the Fe $K\alpha$ 
line. The Laor line energy comes out to be similar to that obtained using a 
Gaussian model. However other Laor fit parameters like the disc inner 
radius or the power law dependence of emissivity are not constrained by the 
data. We therefore used the simple Gaussian model for the line.
For all other dwells too, single Gaussian model gave good fits to the 
data. The relevant free parameters in the in this model, besides the 
normalizations for the model components, are the line energies and widths of 
the Gaussians, powerlaw photon index and the energy of exponential cutoff. The 
results of the spectral deconvolution are shown in 
Table~\ref{tab:fits}. Also shown in Table~\ref{tab:fits} 
are the values of reduced chi-squares obtained for the best fit model and the 
corresponding 3-50 keV isotropic luminosity for a distance of 10 kpc.
(also see Fig.~\ref{d1-espectrum}).

\begin{figure}
\begin{center}
\includegraphics[height=4.0in]{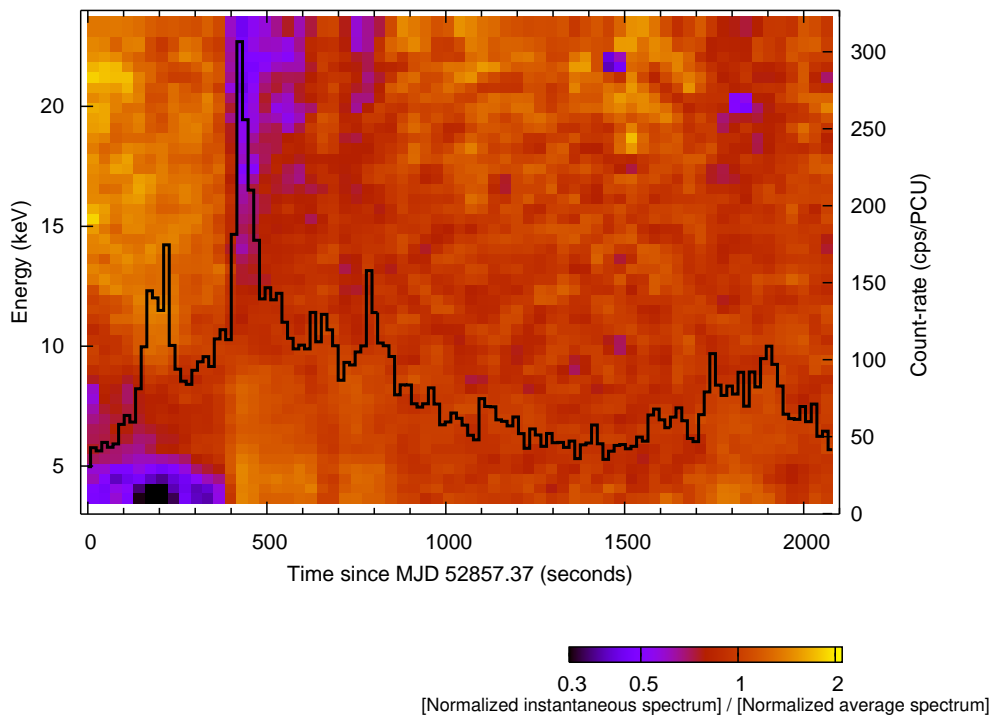}
\caption{
Dynamic Energy Spectrum of V4641 Sgr during dwell (2). Time variation of
normalized spectral energy distribution is shown. For any time bin (each
32s long), the color represents the ratio of flux normalized spectrum
extracted for that time bin divided by a template spectrum. The template
spectrum is the flux normalized, time averaged spectrum of the entire
dwell. Blue-black represents a dip in the spectrum compared with the
template, while yellow represents a bump. The 3-20 keV PCA lightcurve is
overplotted as the solid histogram. The different spectral nature of the
two flares near 200s (hard flare/soft drop-out) and 450s (soft flare/hard
drop-out) are evident.
\label{d2_dyn_espec}}
\end{center}
\end{figure}

\begin{figure}
\begin{center}
\includegraphics[height=4.0in, angle=-90]{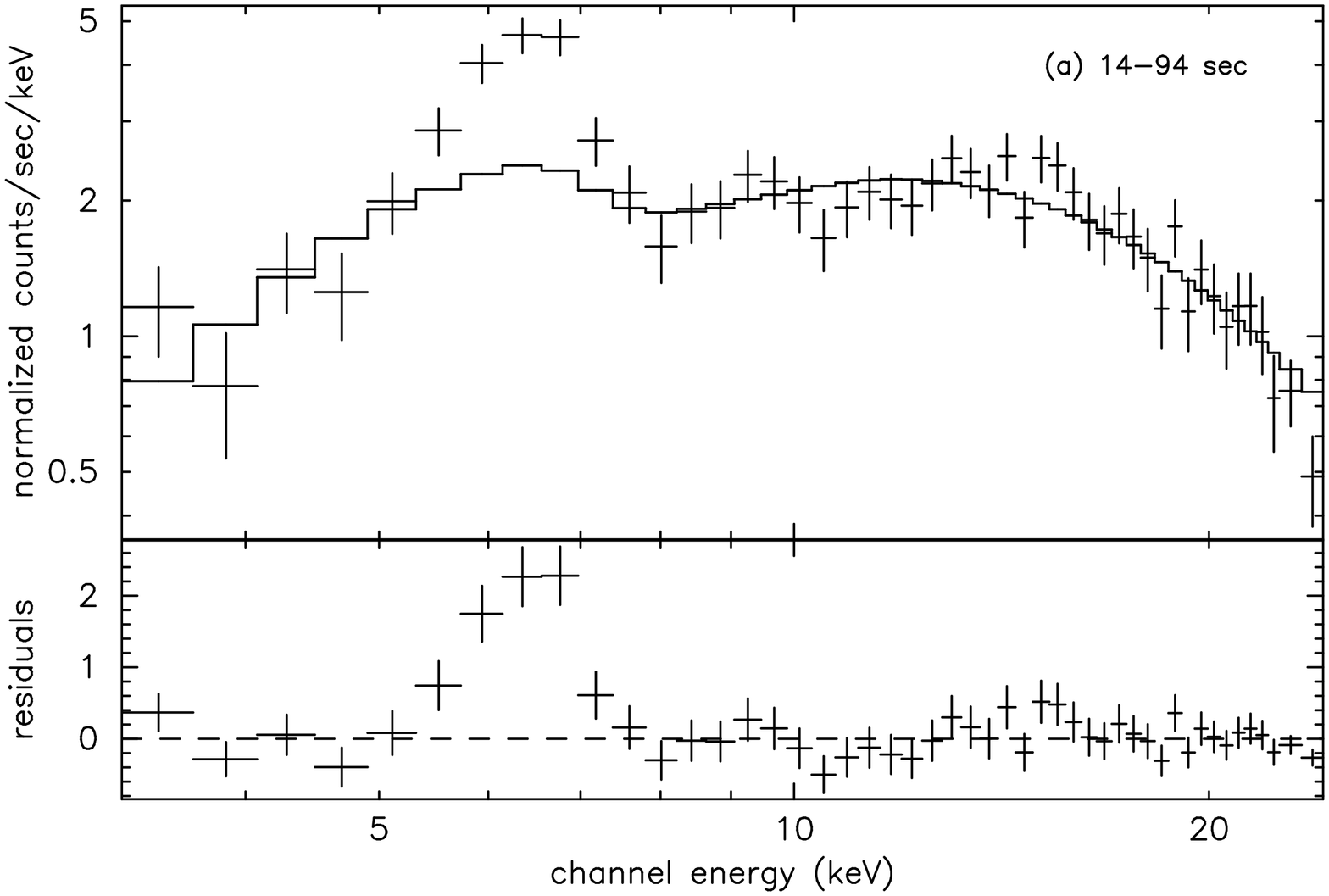}
\includegraphics[height=4.0in, angle=-90]{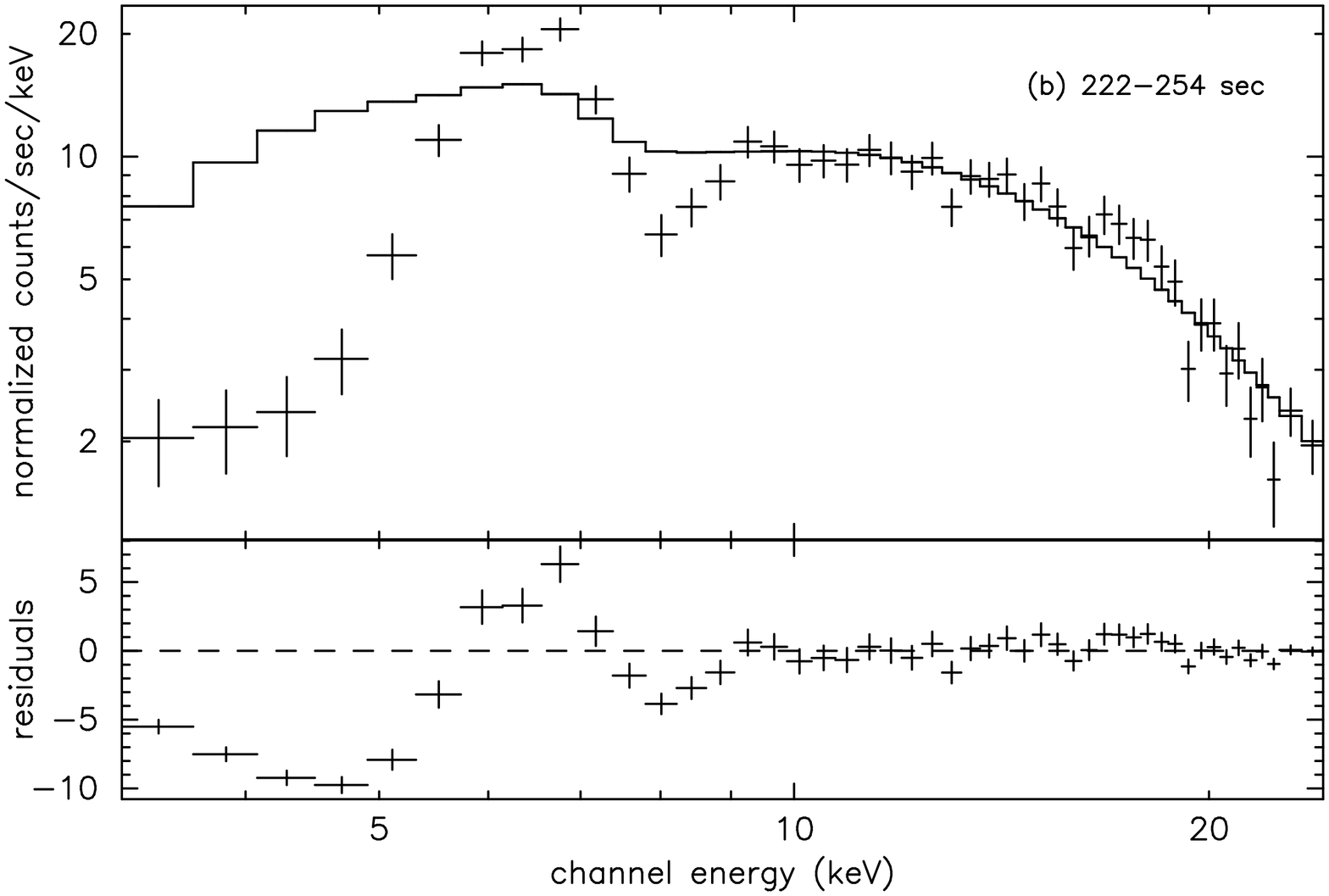}
\caption{Variation of $n_H$ during the hard flare in dwell (2). Panel (a): 
The energy spectrum during 14-94 seconds. Panel (b): The spectrum during 
222-254 seconds when the source was going through a hard flare. The solid 
histogram in both panels is a fit to the 10.0-25.0 keV spectrum with 
$n_H=2.3\times10^{21}$ $atoms/cm^2$. Note that the 
extrapolated fit matches the continuum at lowest energies for panel (a) 
whereas it grossly overestimates the counts during the flare as shown in 
panel (b).\label{nh_spec}}
\end{center}
\end{figure}

Since the broad colors show sharp spectral state changes 
during the second dwell, we have not attempted to create any time averaged 
spectrum of the entire dwell. Instead we present a dynamic energy spectrum 
(DES, Fig.~\ref{d2_dyn_espec}) which shows the spectral evolution of the 
source during this dwell. To create 
the DES we extracted 69 spectra spanning the entire dwell with timing 
resolution of 32 seconds. The spectra were then normalized by count-rate. We 
also extracted the entire time averaged spectrum for the whole dwell and 
normalized by count-rate. This normalized, time averaged spectrum of the 
entire dwell was used as a template
spectrum. Each 32s spectrum was divided by the template spectrum and the 
resulting ratio spectrum constitutes a vertical strip in the DES. The 
color coding corresponds to the ratio of observed spectrum to the template 
spectrum at the time (abscissa) for the energy (ordinate). The overplotted 
solid histogram shows the variation of PCA 3-20 keV lightcurve during the 
dwell. It is evident from the DES that the flares during 200s and 450s are 
intrinsically different in their spectral natures. The first flare near 200s 
is very hard with a sharp dropout of soft photons whereas the second flare
near 450s is essentially soft in nature.
As in the previous dwell, we do not find  any significant blackbody component 
in the spectra. The 3-25 keV spectra are dominated by an iron emission line 
and powerlaw continuum. We modelled the 10-25 keV 
spectra during the first 14-94 seconds of the dwell, i.e. {\em before} the 
hard flare, using the pexrav model. In Fig.~\ref{nh_spec}(a) we show the 
corresponding spectrum and the fit. The solid histogram is the 
Comptonized powerlaw fit to the hard 10-25 keV data which does not account
for the broad Fe $K\alpha$ emission line seen in the data near 6.5 keV. 
Therefore, as expected, when extrapolated to lower energies, the model is 
unable to reproduce the strong iron line near 6.5 keV but the fit matches the 
continuum well at the lowest energies. In sharp contrast to this is the 
spectrum {\em during} the hard flare (222-254 seconds), shown in Fig.~\ref{nh_spec}(b). The solid
histogram, as before, is a Comptonized powerlaw fit to the 10-25 keV 
energy range which is not only unable to reproduce the region near the iron 
line complex, but also largely overestimates the soft photon count-rate in the 
2-5 keV range.

\begin{figure}
\begin{center}
\includegraphics[height=4.0in]{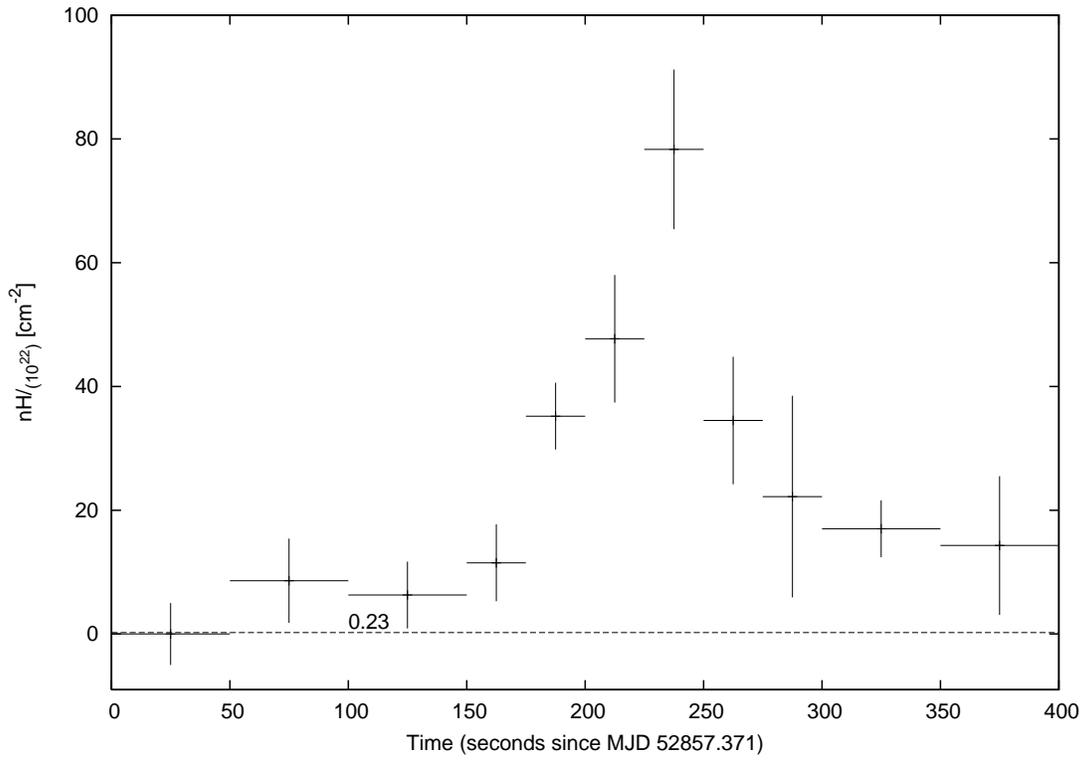}
\caption{Variation of the fit parameter $n_H$ during the hard flare. The 
dashed horizantal line represents the standard adopted value of $n_H$ for 
this source.\label{nh_vary}}
\end{center}
\end{figure}

 In fact, the spectrum during the hard flare 
can not be modelled using any physical model if the column density of the 
absorber is fixed to its standard value of $2.3\times10^{21}$ $atoms/cm^2$
\citep{dl1990}. In contrast, the spectrum before and after the flare can be 
well modelled using this standard column density. We therefore allowed  
the fit parameter $n_H$ which estimates the column density of the warm absorber
to vary during the flare. A comptonized powerlaw \citep{mz1995} was used for 
the continuum and a gaussian for the Fe line. In this model, the variation of
the fit parameter $n_H$, with time, is shown in Fig.~\ref{nh_vary}. Due to low 
count rates and moderate resolution of the spectrometer it is 
difficult to determine precisely the column density of the warm absorber, 
our 90\% confidence range on the estimated maximum colum density during the 
hard flare is $(78.3\pm12.9)\times 10^{22}$ $atoms/cm^2$, almost 2 orders of 
magnitude greater than the rest of the dwell. One possible physical scenario 
that could lead to such an event is the eruption of a 
hard jet followed by an enhanced outflow of mass, as also envisioned by 
\citet{r2002}. Such an outflow moving nearly along the line of sight can cause 
such large changes in observed column density. Other scenarios like variation 
of partial covering fraction cannot be 
ruled out. As regards the strong iron emission line, it is likely that the 
source is obscured and the circumstellar material is excited by hard X-ray 
emission, which in turn produces the line emission with strong equivalent 
widths. 

The 10-50 keV hard energy spectrum for the third dwell could be well described 
by a simple powerlaw ($\chi^2/\nu = 33.7/45$). Unlike the previous dwells, the 
data for the third dwell did not require any Comptonization component. Most likely, 
this is due to the low count-rates and associated larger errorbars which makes 
detection of 
reflection component difficult. At lower energies, the data do not require 
any thermal disc photons. However, the column density of the absorbing 
column was larger than the standard value, during the entire third dwell, which
could be possible if the 
observation was made shortly after a (possibly super-Eddington) mass ejection 
event. An absorption edge near 10 keV in terms of a smeared
edge model \citep{e1994} was also required. The width of the smeared edge was 
fixed to 10keV. 

\begin{figure}
\begin{center}
\includegraphics[height=6.0in, angle=-90]{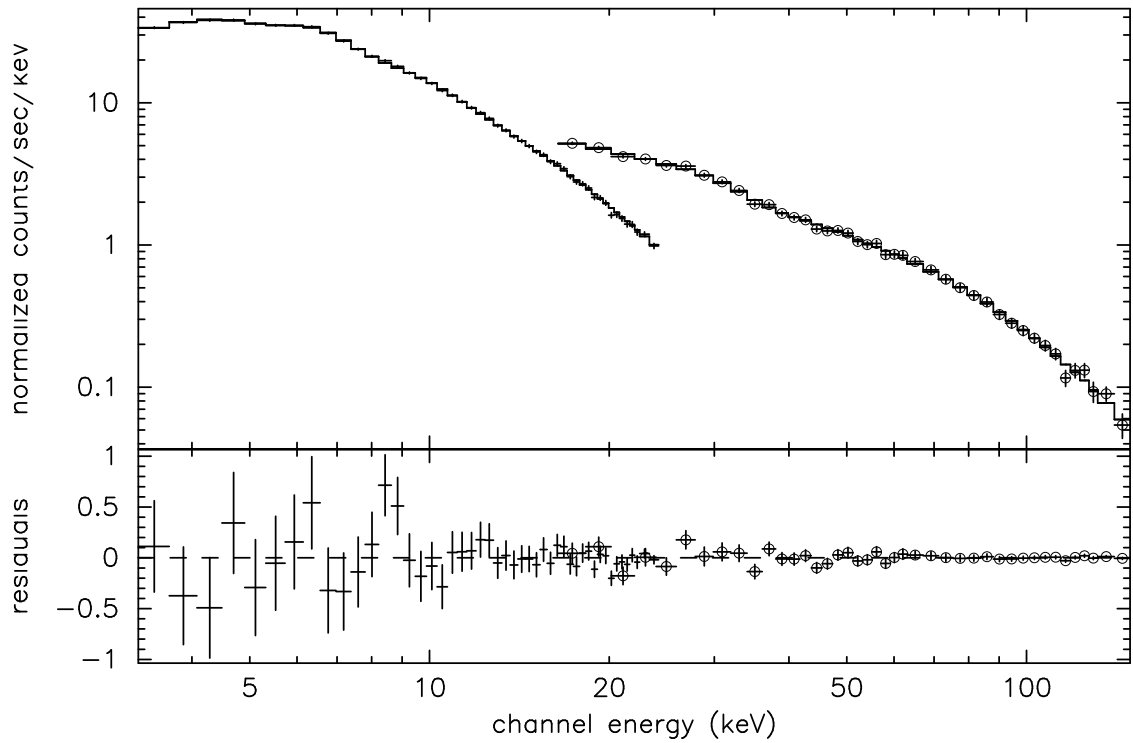}
\caption{Energy spectrum during the first 1000 seconds of dwell (4). The PCA 
data are marked by $+$ and HEXTE data by $\oplus$ symbols. The model consists 
of a Comptonized blackbody and powerlaw to represent the continuum and a 
Gaussian line for the Iron K$\alpha$ emission near 6.5 keV. A warm absorber 
with an $n_H$ of $2.3\times 10^{23}$ atoms/cm$^2$ and a smeared edge near 10 
keV is also included in the model.\label{d4_espec}}
\end{center}
\end{figure}

For the fourth dwell, which was the brightest of all the four pointings, the 
HEXTE spectrum was seen to extend up to 150 keV. In 
Fig.\ref{d4_espec} we show the 3-150 keV PCA+HEXTE spectrum. 
Although the lightcurve showed a variation in flux of an order of magnitude 
over the entire dwell, the X-ray colors or a DES shows that 
the spectral state of the source did not change during first 1000 seconds 
of the dwell. During the 1000-1200 seconds, it appears that the source went 
through another faint hard flare as seen in dwell (2). Unlike the other dwells,
the spectra for this dwell however shows an excess of soft photons, most likely
from an accretion disk. A smeared edge near 10keV was also required. A simple 
model with a Gaussian line
 and two continuum components, viz. a multicolor disk \citep{m1984,ss1973} and 
a Comptonized powerlaw \citep{mz1995} gives a very high temperature of the 
inner edge of the accretion
disk of around 3.3 keV. Assuming a Comptonized blackbody model ({\em compbb}, 
see \citet{n1986} for details) with an electron temperature of the hot plasma 
fixed to 20 keV and a Comptonized powerlaw with fixed photon index of 1.5 and 
cutoff at 150 keV, 
better fits are obtained ($\chi^2/\nu = 75.4/80$, also see Fig.~\ref{d4_espec})
 and lower temperature of the blackbody ($\sim 1.9$ keV). The fitted parameter 
values for the model are given in Table~\ref{tab:fits}. However 
the {\em compbb} model assumes a single temperature blackbody emission as a 
seed photon spectrum whereas in reality, the seed photons most likely are 
comming from an accretion disk with a varying temperature profile. Small 
systematic deviations are seen in the plot of residuals (bottom panel of 
Fig.~\ref{d4_espec}) and could be fitted by adding more models components, but 
due to lack of a thorough understanding of actual undergoing physical processes
 we constrained ourselves to simple models. 

\subsection{Temporal analysis\label{sec:Timing}}

\begin{figure}
\begin{center}
\includegraphics[height=7.0in]{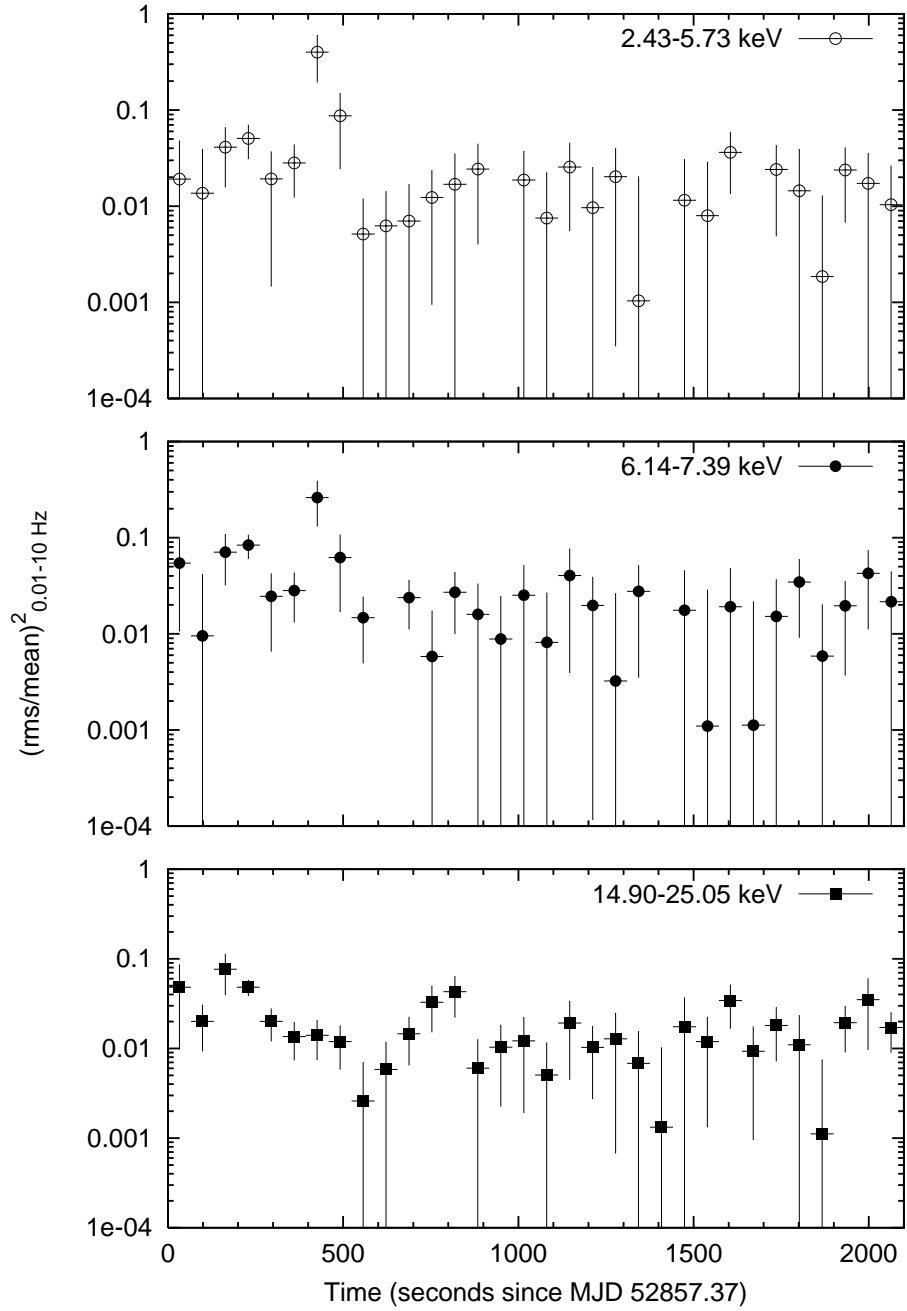}
\caption{0.01-10.0 Hz integrated RMS variability for different energy bands 
during dwell (2) are shown.\label{d2-rms}}
\end{center}
\end{figure}

The rapid state changes seen in the energy spectra during the second dwell 
also left strong signatures in temporal domain.
The rms variability with time, during the second dwell, in the soft (2.4-4.1 
keV), Fe line complex 
(5.3-7.8 keV) and hard (14.9-25.0 keV) energy bands show significant changes 
(Fig~\ref{d2-rms}). The total rms variability in 0.1-10 Hz 
range was calculated for every 64 second long data segments. 
We see that the line variability tracks the soft variability throughout the 
dwell. Therefore it is unlikely that any of these flares is caused solely by 
the variability in the line. Note that the variability increases to a maximum 
near the hard flare (near 200s) for all 
energy bands which may be caused by an outflow crossing the observer's line of 
sight, where the ejected matter has a high rms variability. An enhancement in 
the Comptonizing corona that surrounds the source and gives rise to the high 
variability hard photons (e.g in low/hard states of most compact X-ray 
binaries) could also account for the increase in variability.

Interestingly, we see no change in hard
band variability during the soft flaring event near 450s, but both the soft 
and the line flux variabilities go through a large increase in variability 
during the flare. A dip in the hard band variability is seen after the soft 
flaring event. The hard flaring event could cause instability in the 
circumstellar environment of the source causing a decrease in optical depth 
of the corona and thereby allowing larger number of scattered soft disk 
photons to reach the observer. Thus the soft flare may be due to the 
appearance of a brief `window' in the obscuring circumstellar material, 
allowing 
us a glimpse of the inner regions. The soft-band variability drops just after the soft flare while the hard-band variability slowly increases  
suggesting the rebuilding of the obscuring corona/circumstellar material. After 
the flare around 800s, the source activity decreased and we see no significant
evolution in spectral or temporal domains. 

For the remaining dwells, viz. (1), (3) and (4), the PDS show
little or no rms variability in the frequency range of 0.1-10.0 Hz, in any of 
the soft, Fe line complex or hard (14.9-25.0 keV) energy regimes. This suggests
no temporal variability in 0.1-10 Hz range over the 2.5-25 keV energy range 
during the entire observation. We therefore constructed a white (2.5-25 keV) 
PDS over the entire dwell. 
Below $\sim 1$ Hz the PDS is essentially dominated by 
featureless red noise with little or no signature of any peaked noise. There 
might be some evidence of a flat-topped noise at the lowest frequencies 
($<0.01$ Hz) but it is not statistically significant from the data. Above a few
Hz the spectrum is dominated by Poisson noise.  In 
Table~\ref{tab:fits} we present the rms variability seen in the
2.5-25 keV photons, as a Riemann sum of frequencies between 0.01-10 Hz and the 
powerlaw index ($\alpha$) that characterises the slope of the red noise.

\section{Discussion\label{sec:Summary}}
 Analysis of 
the data presented in this paper suggests that the compact object in V4641 Sgr 
is enshrouded by an optically thick cloud, at least during some periods of its
enhanced activity during the outburst of 2003. Optical 
\citep{c1999} and X-ray \citep{r2002} observations during the previous 
outbursts also suggest similar physical environment.
The cloud could be composed of outflowing matter from the inner regions close 
to the central black hole.
The strong powerlaw dominated flux seen in the observed dwells 
indicate that the source was in the canonical low/hard state. The energy 
spectrum of dwells (1)-(3) do not seem to require a soft disc blackbody 
component.
High column density observed during the outburst can in part cause 
the lack of detection of the soft disk component. High orbital inclination, 
along with presence of gas and dust can also
potentially obscure the accretion disk from the observer and this may be the 
case for V4641 Sgr. In a recent work, \citet{nm2005} have pointed out that the 
X-ray binary systems which have high inclination, show strong variability and 
complex, non-FRED like outbursts. The inner disk, which may be warped and 
combined with the modestly high inclination of the binary orbit, can have an
inclination near $90\degr$. Then small changes in the height of the disk could 
cause rapid obscuration and changes in the observed flux, if the material is 
somewhat thin could cause changes in column density, and if it is thick could 
cause variations in the reflection component.
Although no 
Super-Eddington events were detected in any of the dwells presented in this 
work, given the high variability of the source, it is possible that there were 
short episodes of such Super-Eddington accretion that lead to the formation of 
a dynamic environment around the central engine. 

The color-luminosity diagram for dwell (2) shows that there is significant 
overlap in luminosity between the hard, soft and the non-flaring state. This 
supports the suggestion that the luminosity (and hence the associated inferred 
mass accretion rate, \mdot) is not the only parameter that causes a state 
transition in XRBs \citep{h2001,shs2002,mc2003,mb2004}. The data imply a second
(or more) parameter determining state transition in these sources.

There is evidence for Comptonization, both from the 
presence of a broad Fe line complex near 6.5 keV as well as a characteristic 
Compton hump in $>10$ keV range. The Compton reflection fraction is not well 
constrained by the models which might be due to modification of the emergent 
spectra by an outflow around the source. \citet{ts2005} have recently shown 
that in a relatively cold outflow of $T\about 10^6$ K, emerging photons are 
predominantly downscattered which can lead to an accumulation of excess photons
$\about10$ keV. Such excess \about10 keV was seen for V4641 Sgr and therefore 
strengthens the enshrouded source with an outflow model for this source. Even 
during periods of high luminosity, a simple multi-color disk fit gives an
unrealistically high disc temperature. One possible interpretation of such high
fitted disk temperatures is an existence of a hot electron cloud very near to 
the disk which Compton upscatters the disk photons and converts a significant 
fraction of the disk emission into higher energies as also observed in another 
black hole binary with superluminal jet, XTE J1550-564 \citep{km2004}. The 
origin of broad iron emission lines with equivalent widths up to 1 keV seen 
during the observations could either be Compton reflection from an accretion 
disk or a corona. From the data we were unable to differentiate between the 
origin of the lines.  However since other evidence favor an enshrouded source, 
the coronal line formation scenario seems more likely. Timing analysis shows 
that during all four observations the PSD were dominated by red powerlaw noise 
below few Hz and Poisson noise above it. Besides the powerlaw continuum or the 
constant Poisson level, no other features like breaks or QPOs were seen in any 
of the power spectra. The absence of any signal at frequencies higher than few 
Hertz and a featureless red noise at lower frequencies further support an 
enshrouded obscured source where all the high-frequency/short timescale events
have been smeared out as the radiation passes through the obscuring material.

DM would like to thank Jeroen Homan and Paolo Coppi for many useful discussions.
We would also like to thank the {\em RXTE} team for swiftly triggering our 
target of opportunity observations and the SMARTS observers J. Espinoza and 
D. Gonzalez for taking the data and R. Winnick, who accommodated our many 
requests to revise the observing schedule. This work was supported by National 
Science Foundation grant AST 00-98421, AST 04-07063, NASA ADP grant NAG5-13336 
and data analysis grant NAG5-13777. This research has made use of data obtained
from the High Energy Astrophysics Science Archive Research Center (HEASARC), 
provided by NASA's Goddard Space Flight Center.

\clearpage

\begin{deluxetable}{ccccc}
\setlength{\tabcolsep}{0.02in}
\tabletypesize{\footnotesize}
\tablecolumns{5}
\tablewidth{0pc}
\tablecaption{Observation Log \label{tab:obslog}}
\tablehead{
\colhead{Serial Number} & \colhead{ObsID} & \colhead{Calendar Date} & \colhead{Time}\tablenotemark{a} & \colhead{Exposure}\tablenotemark{b} \\
\colhead{} & \colhead{} & \colhead{(dd-mm-yyyy)} & \colhead{(MJD)} & \colhead{(seconds)} 
}

\startdata
1 & 80054-08-01-00 & 05-08-2003 & 52856.06 & 992  \\
2 & 80054-08-01-01 & 06-08-2003 & 52857.37 & 2192 \\
3 & 80054-08-02-00 & 07-08-2003 & 52858.50 & 944  \\
4 & 80054-08-02-01 & 07-08-2003 & 52858.57 & 1296 \\
\enddata

\tablenotetext{a}{Start of good time interval.}
\tablenotetext{b}{Total exposure time after screening for good time intervals.}
\end{deluxetable}


\begin{deluxetable}{cclllccc}
\rotate
\setlength{\tabcolsep}{0.02in}
\tabletypesize{\footnotesize}
\tablecolumns{8}
\tablewidth{0pc}
\tablecaption{Important Spectral and Temporal Fit Parameters\label{tab:fits}}
\tablehead{
\colhead{} &\multicolumn{4}{c}{SPECTRAL PROPERTIES} &\colhead{} &\multicolumn{2}{c}{TEMPORAL PROPERTIES} \\
\cline{2-5} \cline{7-8} \\
\colhead{Dwell} &\colhead{Model} & \colhead{Component} & \colhead{Parameter} &\colhead{Fitted value} &\colhead{} &\colhead{Red-noise Slope} &\colhead{RMS(\%)}\\ 
}
\startdata
(1) & WABS$\times$ & Gaussian(1) &  Line Energy (keV)	& $6.79^{+0.29}_{-0.24}$ & & & \\
 & (GAUSSIAN+GAUSSIAN+PEXRAV) &    & Line width ($\sigma$, keV)	& $1.48^{+0.24}_{-0.18}$ & & &  \\
 & 	&    & Equivalent width	(keV)		& 0.85  & & & \\
 & & Gaussian(2) &  Line Energy (keV)		& $6.49^{+0.04}_{-0.04}$  & & & \\
 & 	&    & Line width ($\sigma$, keV)	& $0.24^{+0.11}_{-0.21}$  & & & \\
 & 	&    & Equivalent width	(keV)		& 1.06  & & & \\
 & 	& Pexrav   & Photon Index		& $1.01^{+0.15}_{-0.12}$  & & & \\
 &	&    & Cutoff energy (keV)		& $24.0^{+3.4}_{-3.1}$  & & & \\
 & $L_{3-50 keV}=1.5\times10^{37}(d/10kpc)^2$ ergs/s & & \c2=49/55 &  & & $-1.92^{+0.27}_{-0.25}$ & 15 \\
\tableline
(3) & WABS$\times$ & Wabs & $n_H$ ($\times10^{22}$ atoms/cm$^2$) & $13.9^{+2.6}_{-2.4}$  & & & \\
 & SMEDGE$\times$(GAUSSIAN+POWERLAW) & Smedge	& Threshold energy (keV)	& $10.5^{+0.7}_{-0.7}$  & & & \\
 & &    	& $\tau$			& $1.78^{+0.58}_{-0.55}$  & & & \\
 & & Gaussian	& Line energy (keV)		& $6.16^{+0.13}_{-0.13}$  & & & \\
 & &	        & Line width ($\sigma$, keV)	& $0.66^{+0.21}_{-0.22}$  & & & \\
 & &    	& Equivalent width (eV)		& 866  & & & \\
 & & Powerlaw   & Photon index			& $1.64^{+0.14}_{-0.13}$  & & & \\
 & $L_{3-50 keV}=6.4\times10^{36}(d/10kpc)^2$ ergs/s & & \c2=46/56 & & & $-1.36^{+0.15}_{-0.15}$ & 24\\
\tableline
(4) & WABS$\times$ & Smedge & Threshold energy (keV) & $9.98^{+0.24}_{-0.22}$  & & & \\
(0-1000s) & SMEDGE$\times$(COMPBB+PEXRAV & Compbb	& kT (keV)		& $1.96^{+0.37}_{-0.30}$  & & & \\
 & +GAUSSIAN) & 		& $\tau$		& $1.48^{+0.33}_{-0.44}$  & & & \\
 & & Gaussian	& Line energy (keV)	& $6.23^{+0.35}_{-0.35}$  & & & \\
 & &	        & Line width ($\sigma$, keV)	& $0.76^{+0.09}_{-0.08}$  & & & \\
 & &    	& Equivalent width (eV)		& 858  & & & \\
 & $L_{3-50 keV}=6.3\times10^{37}(d/10kpc)^2$ ergs/s & & \c2=75/80 & & & $-1.46^{+0.04}_{-0.07}$ & 72 \\
\enddata
\end{deluxetable}
\clearpage

\end{document}